\documentclass[twocolumn,showpacs]{revtex4}

\usepackage{graphicx}
\usepackage{dcolumn}
\usepackage{amsmath}

\begin{document}

\title{Madelung Energy of the Valence Skipping Compound BaBiO$_3$}
\author{Izumi Hase and Takashi Yanagisawa}
\affiliation{Condensed-Matter Physics Group, Nanoelectronics Research Institute,
National Institute of Advanced Industrial Science and Technology, Tsukuba 
Central 2,
1-1-1 Umezono, Tsukuba 305-8568, Japan\\
CREST, Japan Science and Technology Agency (JST)}

\date{}

\begin{abstract}
Several elements show valence skip fluctuation, for instance, Tl 
forms the compounds
in valence states +1 and +3, and Bi forms in +3 and +5 states.  This kind 
of
fluctuation gives rise to a negative effective attractive interaction and 
the
Kondo effect.  In the compounds of valence skipping elements, the carrier 
doping will
induce superconductivity with high critical temperature.
For example, Ba$_{1-x}$K$_x$BiO$_3$ shows high $T_c$ which is unlikely 
from the
conventional electron-phonon mechanism.  The reason for the missing of 
some valence
states in such valence skip compounds remains a mystery.
We have performed the evaluation of the Madelung potential for BaBiO$_3$, 
and
have shown for the first time that charge-ordered state is stabilized if 
we take into account
the polarization of the oxygen charge. We argue that the effective 
Coulomb interaction energy $U$ may be negative evaluating the local 
excitation energy.

\end{abstract}



\maketitle

In the periodic table more than ten elements show valence skip in all the
compounds they form.
Tl and In form the compounds with the valence states +1 and +3, Bi and Sb 
form with +3 
and +5 states, and Pb and Sn with +2 and +4 valence states.
It is reasonable to expect a new microscopic physics behind this 
phenomenon.
The reason for missing valence states for these elements still remains a 
mystery\cite{var88,and75,har06}.
The valence skipping leads to the Hubbard type model with the negative-$U$ 
attractive 
interaction.
Since the attractive interaction possibly promotes superconductivity, the 
carrier doping 
may induce superconductivity with relatively high $T_c$.  For example, 
Ba$_{1-x}$K$_x$BiO$_3$ shows high $T_c$ which is unlikely from the 
conventional
electron-phonon interaction.
A substitution of Ba with K for the compound BaBiO$_3$ provides 
superconductivity
with $T_c$ exceeding 30K\cite{cha85,mat88,cav88,hin88,sat89}.  
Ba$_{1-x}$K$_x$BiO$_3$ is the perovskite 
three-dimensional superconductor with clear s-wave gap.  
The parent compound BaBiO$_3$ is an insulator
though BaBiO$_3$ should be a metal according to the band 
theory\cite{sle75,cha88}.A band-calculation within the local-density approximation (LDA) leads to a (semi)metallic ground state of BaBiO$_3$\cite{mat83},even under the lattice distortion found in the neutron-diffraction experiments.
The average formal valence
state of Bi is +4.  The insulating state of BaBiO$_3$ would have spin 1/2 
at Bi site if Bi 
with +4 were allowed.  BaBiO$_3$ is, however, the insulator with a charge 
density wave (CDW) 
gap of the order of 2eV.
The valence states +3 and +5 form the CDW state with average charge +4.
The substitution of K for Ba induces carriers as doped holes leading to
superconductivity, which is
reminiscent of the phase diagram of cuprate high temperature 
superconductors.
Since the high critical
temperature found in Ba$_{1-x}$K$_x$BiO$_3$ may be unlikely from the 
conventional electron
-phonon mechanism, one can expect that the negative-$U$ attractive 
interaction would 
cooperate in promoting superconductivity\cite{tin80,sch89,mal91}.  

The valence skip is important in the study of a charge Kondo 
effect\cite{tar91} as well as
valence-fluctuation induced superconductivity\cite{ohn00}.
A superconductor Pb$_{1-x}$Tl$_x$Te with $T_c=1.4$K\cite{mat05} is 
reasonably expected to
belong to the same category.  The mother compound PbTe is a small gap 
semiconductor.
For Tl concentrations up to the solubility limit 1.5 percent, the material
shows superconductivity 
with a remarkably high $T_c$ as a low carrier metal.  Tl impurities as 
negative-$U$ centers
could leads to superconductivity and at the same time Kondo effect as a 
charge 
analog\cite{tar91,dze05}.

The purpose of this paper is to investigate the mystery of missing 
valence  states
evaluating the intra-atomic energy for Bi$^{4+}$ ion.
In general, the total energy is the sum of the one-body potential energy 
and the
interaction energy.  The Coulomb interaction is primarily important for the
insulator BaBiO$_3$.
The curvature of the total energy as a function of the valence $n$, which 
is
closely related with the intra-atomic energy $U$, may be presumably 
changed due
to the Coulomb interaction.
In the way stated we calculate the Madelung energy of BaBiO$_3$.  
Let $q_i$ be the charge on the site $i$, the total Coulomb energy is
\begin{equation}
E= \frac{1}{2}\sum_{i\neq j}\frac{q_iq_j}{r_{ij}},
\end{equation}
where $r_{ij}$ is the distance between the sites $i$ and $j$.
If we define the potential at
the site $i$ as
\begin{equation}
V_i= \sum_{j\neq i}\frac{q_j}{r_{ij}},
\end{equation}
the total Coulomb
energy is
\begin{equation}
E= \frac{1}{2}\sum_iq_iV_i.
\end{equation}
The potential $V_i$,
called the Madelung potential, is evaluated using the Ewald 
method.\cite{kit05,kon89}
We calculated the Madelung potential and the total energy for the valence
skipping compound BaBiO$_3$.
We assign the formal charges as
\begin{equation}
(A)~~{\rm Ba}:2+,~{\rm Bi}:4+,~{\rm O}:2-
\end{equation}
\begin{equation}
(B)~~{\rm Ba}:2+,~{\rm Bi(1)}:3+,~{\rm Bi(2)}:5+,~{\rm O}:2-
\end{equation}
where in the latter case the charge density wave with valences $3+$ and 
$5+$
is assumed.
Let us denote the number of Ba atoms as $N_A$, then the number of Bi and O 
atoms are
$N_A$ and $3N_A$, respectively.
The total Madelung energy for the case (A) is
\begin{eqnarray}
E_{Madel}&=&
\frac{1}{2}(N_Aq_{Ba}V_{Ba}+N_Aq_{Bi}V_{Bi}+3N_Aq_OV_O)\nonumber\\
&=& \frac{N_A}{2}(q_{Ba}V_{Ba}
+q_{Bi}V_{Bi}+3q_OV_O),
\end{eqnarray}
where $q_{Ba}=2$, $q_{Bi}=4$ and $q_O=-2$.
For the case
(B) the total Madelung energy is
\begin{eqnarray}
E_{Madel}&=& \frac{N_A}{2}(q_{Ba}V_{Ba}+\frac{1}{2}
q_{Bi(1)}V_{Bi(1)}+\frac{1}{2}q_{Bi(2)}V_{Bi(2)}\nonumber\\
&+& 3q_OV_O),
\end{eqnarray}
for $q_{Ba}=2$,
$q_{Bi(1)}=3$, $q_{Bi(2)}=5$ and $q_O=-2$.
The intra-atomic repulsion energy $U$ can be calculated if the total energy
is obtained.  For the valence $n$, the energy $U$ is defined as
\begin{equation}
U_n= E_{n+1}+E_{n-1}-2E_n,
\label{udef}
\end{equation}
where $E_i$ is the energy of the valence $i$ state.
The $U$ for Bi$^{4+}$ per two bismuth atoms has been estimated from the 
measured 
ionization
energies of the elements as\cite{var88}
\begin{equation}
U_{4+}= E_{ion}(Bi^{3+})+E_{ion}(Bi^{5+})
-2E_{ion}(Bi^{4+})=10.7{\rm eV}.
\end{equation}
Here $E_{ion}(Bi^{n+})$ is the ionization energy of the element Bi$^{n+}$.
This estimated value is quite large although it is smaller than
for Bi$^{3+}$ ($U_{3+}=19.7$eV) and Bi$^{5+}$ ($U_{5+}=32.3$eV).
Thus the simple estimate results in the positive $U$.

We compare the total energy of the uniform state (A) and the ordered CDW 
state (B), where 
the total energy is the sum of the total Madelung energy and the one-body 
(ionic)
potential energy .
The total energy difference is
\begin{eqnarray}
\Delta{} E&=&E(Ba_2^{2+}Bi^{3+}Bi^{5+}O_6^{2-})
-E(Ba_2^{2+}Bi^{4+}Bi^{4+}O_6^{2-})\nonumber\\
&+& E_{ion}(Bi^{5+})
+ E_{ion}(Bi^{3+})-2E_{ion}(Bi^{4+}),
\end{eqnarray}
where $E(Ba_2^{2+}Bi^{3+}Bi^{5+}O_6^{2-})$
and $E(Ba_2^{2+}Bi^{4+}Bi^{4+}O_6^{2-})$ are the total Madelung energies 
per two
BaBiO$_3$'s, respectively.
The charge-ordered state is stabilized if $\Delta E$ is negative.
The Madelung potential $V_a$ ($a$=Ba,Bi and O) are shown in the Table I
where we use the convention that the potential $V_i$ is negative for the 
positive charge $q_i>0$.
The total potential energy per BaBiO$_3$, $E_{tot}/N_A$, is -81.95 eV for 
(A)
and -83.40 eV for (B),
respectively, which are shown in the first and second rows of the Table I.
Apparently the case (B)
with the charge density wave has the lower energy. From these values, we 
obtain 
$\Delta E=7.8$eV which is still positive. It is now important to notice 
that 
the $\Delta E$ is reduced if we consider the movement of the oxygen ions. 
The neutron diffraction demonstrated that the Bi ions occupy two 
crystallographically
inequivalent sites with different Bi-O bond lengths\cite{cox76}.
It has been reported by the neutron diffraction experiment that 
the oxygen atoms are out of the center of two Bi ions  by 4
percent of the Bi-O bond length due to lattice 
relaxation\cite{cha88} (see Fig.1).
Thus it is reasonable to move the positions of oxygen atoms with different 
Bi-O bond
lengths in calculating the 
Madelung potential.
In order to take into account the polarization of electrons of oxygen 
atoms, we have further
moved the center of point charges more than reported by the neutron 
diffraction.
It is plausible to assume that the center of the negative charges is more 
close to the 
neighboring
positive charges Bi$^{5+}$ than the nucleus of the atom.
The evaluated results are shown in the Table I,
and are also shown in Fig.2 as a function of the bond length. We found
that the stabilization energy $\Delta{} E$ is reduced linearly as  a 
function of the bond
length O-Bi(1)
and becomes negative for about 10 percent movement.
This indeed shows the stability of the observed CDW state.

Moreover, chemical bonding between Bi and O ions can further stabilize
 this CDW state. Bi$^{5+}$ ion can gain energy as the neighboring oxygen ions move 
toward it, lowering the energy of the oxygen states while raising the energy of the empty Bi s states, as compared to the Bi$^{3+}$ where those s-states are occupied. We consider Bi$^{5+}$ O$^{2-}_{6}$ cluster and the three-parameter model shown by Mattheiss and Hamann\cite{mat83} for simplicity. Adopting more precise five-parameter model does not change the result seriously. The relevant orbitals are $|Bi-s>$ and $|O-A_{1g}>=(|p_{1}>+|p_{2}>+|p_{3}>+|p_{4}>+|p_{5}>+|p_{6}>)/\sqrt{6}$ orbitals, where $|p_{n}> (n=1,..,6)$ denotes the $n$-th oxygen p-orbital directing the Bi site. The secular equation is:
\begin{eqnarray}
\begin{vmatrix} \Delta - \epsilon & t \\ t & - \epsilon  \end{vmatrix} &=&0
\label{seceq}
\end{eqnarray}   
here $\Delta $ is the difference of the one-electron orbital energy $\Delta =\epsilon (Bi-s) - \epsilon (O-A_{1g})$, and $t$ is the effective transfer matrix $t=\sqrt{6 }(sp\sigma )$. Using the well-known relation $(sp\sigma ) \propto  d^{-2}$ \cite{harel} and the distortion $x=0.26$, then the energy gain (loss) by the distortion is 0.42eV per electron for the bonding (antibonding) state. Since the Bi$^{5+}$ site has two empty states, thus the total energy gain is 0.84eV per Bi$^{5+}$.On the other hand, Bi$^{3+}$ ion cannot gain energy because there is no empty s-states. This effect apparently promotes forming the CDW state.

In the above discussion we only considered the insulating states. However,
in the uniform BaBiO$_3$ the system may be metallic. In fact, 
band-calculations show that uniform BaBiO$_3$ with no distortion and oxygen 
polarization should be metallic\cite{mat83,tak87}. The width of Bi6s band 
is about 4eV, and the Fermi level is just the middle of this band. Thus 
considering the small density of states at the bottom of this band,
the average energy gain forming Bloch state is less than 1eV per electron. Each 
Bi$^{4+}$ has one Bi6s electron, so that the total energy of the uniform state 
may go down less than 2eV. However, this value is not so large compared with 
the large energy gain of CDW state, and a further but little change of the 
negative charge, namely $x \sim 0.285$, can compensate this effect.

In order to further investigate the possibility of the
{\em negative U} originating from the long-range Coulomb interaction, 
we consider a state excited locally from the CDW ground state:
 a (3+, 5+) Bi pair (e.g. Bi atoms at (0,0) and (0.5,0) in Fig.1) is 
changed into a (4+, 4+) pair in the CDW background.  
The excitation energy for this process is
\begin{eqnarray}
E_{loc}&=&\frac{1}{2}(V_{Bi1}-V_{Bi2})
-\frac{1}{2}\sum_i q_i(\frac{1}{r_i} - 
\frac{1}{|\mathbf{r_i}-\mathbf{r}_{B}|}) - U_{4+}.
\label{eloc}
\end{eqnarray}
Here $\mathbf{r}_{B}=a_{1}\mathbf{e}_{x}$ denotes the position vector of 
Bi(2) at $\mathbf{e}_x=(0.5, 0, 0)$, and $a_{1}$ is twice the average 
Bi-Bi 
distance ($a_1=8.700$\AA{}).  $V_{Bi1}$ and $V_{Bi2}$ are the Madelung 
potential at the Bi(1) and Bi(2) site, respectively.
 The first term in eq.(\ref{eloc}) is the change of the Madelung energy of 
the Bi pair, and the second term denotes that of the Madelung potential of 
the other ions.  The third term shows the energy gain of the ionic 
(one-body potential) energy for (3+, 5+) $\to$ (4+, 4+), shown in eq.(9). 
We performed the three-dimensional summation in real space for sites in 
the sphere of the radius $R$ using the procedure of Harrison\cite{har061}. 
This summation converged well for $R \sim 200a_{0}$ (see Ref.22).  We do 
not suffer from the usual extreme oscillation of the potential in this 
case because we only consider the difference of the Madelung energy.  Thus 
we do not need to add the $-Q/R$ term in contrast to the method in Ref.22. 
In the case of the Bi(1)-O bond-length $x=0.282$, the local excitation 
energy is $E_{loc}=6.92 + 10.56 - 10.7 = 6.78$eV. 
Hence the CDW state is stable against local valence fluctuation. Rice and 
Sneddon examined the low-energy effective Hamiltonian for the CDW state of 
BaBiO$_3$ and insisted that there are two types of 
excitations\cite{ric81}. One is the single-particle excitation mentioned 
above, and the other is the two-particle excitation which is an exchange 
of local charges as (3+, 5+)$\to $ (5+, 3+).  The excitation energy for 
(3+, 5+) $\to $ (5+, 3+) is extremely large because the second term of 
eq.(11) is twice as large as the process (3+, 5+) $\to $ (4+, 4+). 
From the definition of $U_n$ in eq.($\ref{udef}$), the quantity 
$U_{4+}^{local}\equiv -E_{local}$ resembles the intra-atomic energy.
The value $ U_{4+}^{local}=-6.78$eV seems too large as the 'negative' $U$; 
the high dielectric constant and doped carriers, however, may reduce 
$E_{loc}$.

In summary,
we have evaluated the Madelung potential for the valence skipping compound 
BaBiO$_3$ using
the Ewald method.  The total energy of the charge-ordered CDW state was 
calculated using the 
Madelung potential.
We have evaluated the excitation energy for the local disorder (3+, 5+) 
$\to $ (4+, 4+).
It was shown that the polarization of electrons reduces the energy 
difference $\Delta E$ and that an attractive 'negative' $U$ is presumably 
realized if the center of the negative charges is 
moved from the center of the Bi-Bi bond by about 10 percent of
the bond length.  
This indicates a possibility that the high critical temperature of
Ba$_{1-x}$K$_x$BiO$_3$ is due to the valence-skipping induced
negative $U$ as well as the electron-phonon interaction.
The results also suggests that the elements with high electronic 
polarizability have 
strong possibility to induce negative $U$. 
Since the Te$^{2-}$ ion has high polarizability\cite{kit05}, a 
negative $U$
induced superconductivity 
is likely present in Tl doped PbTe.
The valence skip may provide a new idea for the material design of new
superconductors.

We thank K. Yamaji and J. Kondo for stimulating discussions and 
encouragement.

\begin{table}[t]
\caption{Madelung potential in units of eV.
The total electrostatic energy $E_{Mad}/N_A$ and
the stabilization energy of the CDW state $\Delta{} E$ defined as eq.(10) 
are shown in units of eV. Note that $E_{Mad}/N_A$
is for one
BaBiO$_3$.  The bond length between bismuth (1) and oxygen  atoms is 
normalized as
0.25 when there
is no lattice relaxation.}
\label{t1}
\begin{center}
\begin{tabular}{@{\hspace{\tabcolsep}\extracolsep{\fill}}cccccccc}
\hline
Bond length &       & Ba & Bi(1) & Bi(2) & O & $E_{Mad}/N_A$ & $\Delta{} 
E$\\
\hline
0.25  & (A) & -8.917 & -20.487 & -20.487 & 10.685 & -81.95 & \\
0.25  & (B) & -8.917 & -17.594 & -23.379 & 10.685 & -83.39 & 7.8 \\
\hline
0.26  & (A) & -8.908 & -19.321 & -21.789  & 10.661 & -82.00 & \\
0.26  & (B) & -8.908 & -16.428 & -24.681  & 10.867 & -84.68 & 5.3\\
\hline
0.27  & (A) & -8.881 & -18.277 & -23.244 & 10.589 & -82.17 & \\
0.27  & (B) & -8.881 & -15.384 & -26.136 & 11.003 & -86.10 & 2.8 \\
\hline
0.28  & (A) & -8.838 & -17.341 & -24.874 & 10.465 & -82.45 & \\
0.28  & (B) & -8.838 & -14.448 & -27.766 & 11.093 & -87.66 & 0.26\\
\hline
0.282 & (A) & -8.827 & -17.166 & -25.223 & 10.434 & -82.52 & \\
0.282 & (B) & -8.827 & -14.273 & -28.115 & 11.105 & -87.99 & -0.24\\
\hline

\end{tabular}
\end{center}
\medskip
\end{table}

\begin{figure}[tb]
\begin{center}
\includegraphics[width=\columnwidth]{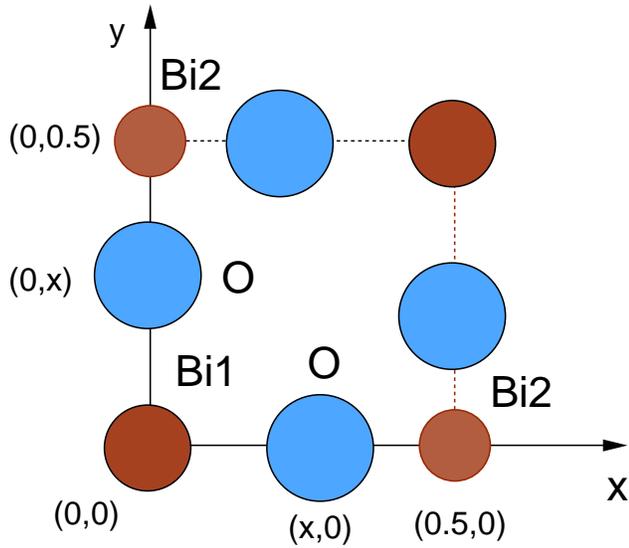}
\end{center}
\caption{The positions of bismuth and oxygen atoms in the plane of $z=0$. 
Here we ignored the small tilting of BiO$_6$ octahedra.}
\label{f1}
\end{figure}

\begin{figure}[tb]
\begin{center}
\includegraphics[width=\columnwidth]{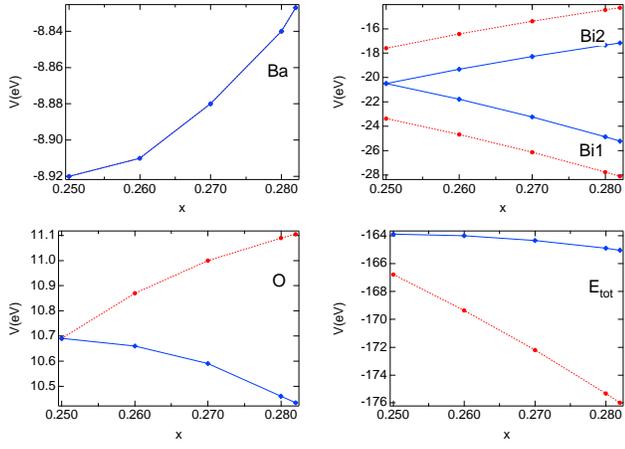}
\end{center}
\caption{The Madelung potential at the Ba, Bi and O sites and the
total energy as a function of the bond length.  
The solid lines and the dotted lines correspond to the case (A) and the 
case (B) (with the
charge density wave), respectively.
The bond length is normalized as 0.25 when there is no
lattice relaxation.}
\label{f2}
\end{figure}

\end{document}